\documentclass{PoS}

\usepackage{subcaption}

\title{Search for Primordial Black Hole evaporations with H.E.S.S}

\ShortTitle{Search for PBH evaporations with H.E.S.S}

\author{\speaker{Thomas Tavernier}\\
        CEA-IRFU, Gif-Sur-Yvette, France\\
        E-mail: \email{thomas.tavernier@cea.fr}
        }
\author{Jean-Fran\c cois Glicenstein\\
        CEA-IRFU, Gif-Sur-Yvette, France\\
        E-mail: \email{jean-francois.glicenstein@cea.fr}}
\author{Fran\c cois Brun\\
        CEA-IRFU, Gif-Sur-Yvette, France\\
        E-mail: \email{francois.brun@cea.fr}}
\author{for the H.E.S.S. Collaboration\footnote{for collaboration list see PoS(ICRC2019)1177}}

\abstract{Following LIGO results of intermediate mass black holes mergers, the idea that dark matter is composed of Primordial Black Boles (PBH), made a recent comeback.
PBHs might be formed in the early Universe through a variety of mechanisms, best known being the gravitational
collapse of overdense regions due to density fluctuations. It's widely accepted that black hole loose there  mass over time through the Hawking radiation process.
Since the particle emission rate increases with black hole temperature, PBH
evaporation is a runaway process that eventually leads to a violent explosion. It has been argued that a class of ultra short gammay ray bursts are actually PBH explosions. The current upper limits on the local PBH explosion rate lie in the $10^4$--$10^5$ $\mathrm{pc}^{-3}$ $\mathrm{yr}^{-1}$ range.

This contribution reports on the search for TeV $\gamma$-ray bursts with a timescale of a few seconds, as expected from the final stage of PBHs evaporation, using 2700 hours of H.E.S.S. extragalactic observations. We present the search algorithm, statistical estimations strategies and results of this analysis.}

\FullConference{36th International Cosmic Ray Conference -ICRC2019-\\
                July 24th - August 1st, 2019\\
                Madison, WI, U.S.A.}

\begin{document}
\section{Introduction}
Primordial black holes (PBH) can form in the early Universe via a variety of mechanisms such as the gravitational collapse of over-dense regions with significant density fluctuations, pressure reduction or bubble collisions during cosmic phase transitions, and collapse of topological defects such as cosmic strings or domain walls. The mass function of PBHs depends on the actual formation mechanism. PBHs could have masses ranging from 10$^{-5}$g for PBHs created at the Planck time up to roughly $1 M_{\odot}$ for PBHs created during the QCD phase transition.    

Black holes 
were predicted by Hawking \cite{1974Natur.248...30H} 
to radiate off particles with a black body spectrum of energies. The emission can be described by an effective temperature
\begin{equation}
 T_{\mathrm{BH}} = \frac{M_p^{2}}{8\pi M_{\mathrm{BH}}},
\label{eq:Tvsmass}
\end{equation}
where $\mathrm{M_p}$ and $M_{\mathrm{BH}}$ are the Planck mass and the PBH mass respectively. 
Black holes lose their mass by Hawking radiation at a rate inversely proportional to their squared mass.

The evaporation rate is thus an increasing function of temperature by equation
(\ref{eq:Tvsmass}). It also depends on the particle physics model at high temperatures \cite{1991Natur.353..807H}.
Since the particle emission rate increases with black hole temperature, PBH evaporation is a runaway process that eventually leads to a violent explosion and bursts of particles. 
PBHs whose initial mass does not exceed 5 $\times$ 10$^{14}$g are expected to have fully evaporated within the 10$^{10}$ years of our Universe history. Consequently, PBHs a little more massive than this will still be emitting particles at a rate large enough so that they would be detectable.\\

The best method for constraining the density of PBHs is through their $\gamma$-ray emission since it allows to integrate over the whole evaporation history. Previous searches have attempted to detect a diffuse photon signal from a distribution of PBHs 
or to search directly for the final stage emission of an individual hole 
The search for direct PBH explosions through $\gamma$-ray bursts did not find any evidence of their presence yet. The current upper limits on the local PBH explosion rate lie in the 10$^{3}$-10$^{6}$ pc$^{-3}$ yr$^{-1}$ range. Searches via anti-protons emission have also been carried out. Since the PBH anti-protons spectrum should present a distinct signature compared to the secondary anti-protons spectrum, the measured cosmic anti-protons flux at earth has been fitted by several experiments in a PBH scenario. Limits as good as those derived in the search for a PBH diffuse $\gamma$-ray background have been obtained. 

The present contribution reports on the search for TeV $\gamma$-ray bursts with a timescale of a few seconds, as expected from the final stage of PBHs evaporation, using the H.E.S.S. array of Imaging Atmospheric Cherenkov Telescopes (IACTs).

The modelling of the PBH signal and the search strategy are described in Sec. \ref{sec:modeling}. The H.E.S.S. data selection and processing is then presented in Sec. \ref{sec:observations}. The burst search algorithm, background estimation and time window choice are discussed in section \ref{sec:analysis}. Finally, the results on the local explosion rate, and the comparison to existing and future limits are given in \ref{sec:results}.

\section{Predictions for the PBH evaporation signal}\label{sec:modeling}

The  instantaneous $\gamma$-ray spectrum d$^{2}$N/dE$_{\mathrm{\gamma}}$dt 
emitted in its last seconds by the PBH depends on the elementary particle 
mass spectrum and thus on the particle physics model. 
It is assumed in this paper 
that the standard model of particle 
physics remains valid at high ($> 200 GeV$) temperatures.  
Possible PBH atmosphere effects on the evaporation signal are neglected.

In the framework of the standard model of particle physics, the integrated 
spectrum of photons emitted during the time $\Delta$t before total evaporation of the PBH
and observed by a detector distant of $r_0$ is
is given by 
\begin{equation}
N(>E)=
0.22 \left(\frac{0.1 \mathrm{pc}}{r_0}\right)^2
(\frac{GeV}{Q})^{2}(5/14(\frac{E}{Q})^{3/2}+3(\frac{E}{Q})^{1/2}+5/6(\frac{Q}{E})^{1/2}-5/3(\frac{E}{Q})-5/2+1/150)
\end{equation}
for $E < Q$ and
\begin{equation}
N(>E)=
0.22 \left(\frac{0.1 \mathrm{pc}}{r_0}\right)^2
(\frac{GeV}{E})^{2}(1/42+1/150)
\end{equation}
for $E \ge Q.$
In the above equations, E is the energy, and Q relates to $\Delta$t by 
$Q=40 \mbox{TeV} (1 \mbox{s}/\Delta \mbox{t})^{1/3}.$

The theoretical 
average 
number of $\gamma$-rays emitted from a PBH located at a distance r and in the direction ($\mathrm{\alpha}$,$\mathrm{\delta}$) in the sky, during the last $\mathrm{\Delta}$t seconds of its life is given by:
\begin{equation}
N_{\mathrm{\gamma}}(r,\mathrm{\alpha},\mathrm{\delta},\mathrm{\Delta}t) = \frac{1}{4\mathrm{\pi}r^{2}} \int_{0}^{\mathrm{\Delta}t}dt \int_{0}^{\infty} dE_{\mathrm{\gamma}} \frac{d^{2}N}{dE_{\mathrm{\gamma}}dt}(E_{\mathrm{\gamma}},t) A(E_{\mathrm{\gamma}},\mathrm{\alpha},\mathrm{\delta}),
\end{equation}
where d$^{2}$N/dE$_{\mathrm{\gamma}}$dt is the instantaneous $\gamma$-ray spectrum emitted by the PBH at a time t before complete 
evaporation. 
The photon spectrum emitted during a PBH explosion becomes harder in the very last seconds of the PBH life. The signature of a PBH explosion is thus a short, few seconds long, burst of high energy photons.
 
The evaporation photon spectrum has to be folded with the H.E.S.S. acceptance A(E$_{\mathrm{\gamma}}$,$\mathrm{\alpha},\mathrm{\delta}$) to take into 
account the instrument's efficiency in collecting $\gamma$-rays of energy E$_{\mathrm{\gamma}}$ at equatorial coordinates $(\mathrm{\alpha},\mathrm{\delta})$ in the sky. The response of the H.E.S.S. instrument to $\gamma$ rays depends on the 
zenith angle and offset angle of observation.  
The acceptance $ A(E_{\mathrm{\gamma}},\mathrm{\alpha},\mathrm{\delta})$ is an average over many runs with different zenith and offset angle. 
%

The probability of detecting a burst of size b when observing a PBH which emits N$_{\mathrm{\gamma}}$(r,$\mathrm{\delta}$,$\mathrm{\alpha}$,$\mathrm{\Delta}$t) $\gamma$-rays follows the Poisson statistics:
\begin{equation}
P(b,N_{\mathrm{\gamma}}) = e^{-N_{\mathrm{\gamma}}} \frac{N_{\mathrm{\gamma}}^{b}}{b!}
\end{equation}
Integrating this probability over space, and summing over each run gives the number of expected bursts of size b to be detected in the data:
\begin{equation}
n_{sig}(b,\mathrm{\Delta}t) = \mathrm{\dot{\rho}_{PBH}}V_{\mbox{eff}}(b,\mathrm{\Delta}t)
\label{eq:ngammath}
\end{equation} 
where $\dot{\rho}_\mathrm{PBH}$ is the local PBH explosion rate and the effective space-time volume of PBH detection is defined by 
\begin{equation}
V_{\mbox{eff}}(b,\mathrm{\Delta}t) =
\sum_{i} T_{i} \int d\mathrm{\Omega_{i}} \int_{0}^{\infty} dr r^{2} P_{i}(b,N_{\mathrm{\gamma}}),
\label{eq:effectivevolume}
\end{equation}
where the index i goes over each run of the H.E.S.S. dataset, T$_i$ and d$\mathrm{\Omega_i}$ being the corresponding run live time and observation solid angle respectively.\\

The effective volume can be written explicitely as
\begin{equation}
    V_{\mbox{eff}}(b,\mathrm{\Delta}t) =
    \sum_{i} T_{i} \mathrm{\Omega_{i}}
    \frac{{(r_0 \sqrt{N_0})}^3}{2} \frac{ \Gamma(b-3/2) }{\Gamma(b+1) }
\end{equation}
where $N_0$ is the observed number of photons from a PBH at $r_0.$

\section{H.E.S.S. observations and data reduction}\label{sec:observations}

H.E.S.S. is an array of five imaging atmospheric Cherenkov telescopes dedicated to observing very-high energy (VHE) $\gamma$-rays with energies above $50$~GeV 
from astrophysical sources. It is located in the Khomas Highland of Namibia at an altitude of 1800 m above sea level. The first four telescopes (12m diameter) have been installed in 2003 (H.E.S.S-1 phase of the experiment) and have been operational since 2004 . A fifth telescope (28m diameter), installed at the center of the original array started its operations in 2012. The analysis described here uses only data collected with the four telescopes of H.E.S.S-1.
In this configuration, the system has a field of view of $5^{\circ}$ in diameter. It allows to reconstruct the incoming direction of the primary gamma-ray with an accuracy of $\sim 0.08^{\circ}$ and its energy with a resolution of $15\%$. 

\begin{figure}[htb]
\centering{\includegraphics[scale=0.5, clip=true]{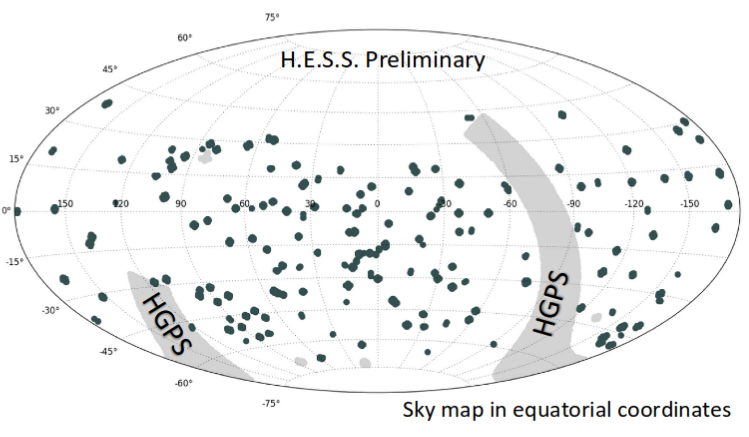}}
\caption{Sky coverage of the H.E.G.S}
\label{fig:hegs}
\end{figure}

The data used for this analysis are all the H.E.S.S.-1 observations taken between January 2004 and January 2013 towards sky position with Galactic latitudes $|b| > 10^{\circ}$. One H.E.S.S. observation run consists of data taken towards the same position on the sky during $\sim 28$ minutes. Some regions of the sky (LMC, SMC, region of SN 1006) were excluded from this analysis as well as runs of poor quality, affected for instance by bad weather or technical problems. The data set, the H.E.S.S. Extra Galactic Survey (H.E.G.S., see Fig \ref{fig:hegs})
comprises 6240 runs, corresponding to more than 2600 hours of observations. 
The Model analysis \cite{deNaurois2009} is applied to all the runs in order to suppress the background of hadronic cosmic rays and reconstruct the direction and energy of 
the gamma-ray candidates. The arrival times of these so-called ``gamma-like'' events are extracted together with their reconstructed parameters. For each run, gamma-like events with a distance to the center of the camera larger than 2 degree are excluded.


\section{Analysis}\label{sec:analysis}

The principle of the analysis is counting the number of clusters of few photons (2 to 10) which arrive in coincidence in a short time-scale and
with directions compatible with the H.E.S.S. instrument PSF.
The clusters are found with the open source implementation\cite{scikit-learn} of the DBSCAN 
(density-based spatial clustering of applications with noise) algorithm \cite{DBSCAN}. 
Since the algorithm of DBSCAN finds all nearest neighbors, the photon clusters may be larger than the instrument PSF and last onger than $\Delta t.$ 
As explained in Section \ref{sec:dbscan}, the clusters 
found by DBSCAN are reprocessed to remove all photons of the cluster which are outside the space or time limits.

In H.E.S.S. observations, photons will be associated in clusters
by chance association. 
For each run, this false positive background is estimated running the same analysis on the same data-set  with randomized photon arrival times. The evaluation of the background is described in Section \ref{sec:bckgrd}.  

\subsection{Cluster counting algorithm}\label{sec:dbscan}

The DBSCAN algorithm  requires two parameters: a minimum distance ($\epsilon$) and a minimum number of points required to form a cluster (minPts). 
Starting with an arbitrary unvisited 
point, when $n>\mathrm{minPts}$ photon arrival events are found in its $\epsilon$-neighborhood, a cluster is created and built step by step. For our analysis we use a minimum number of points $\mathrm{minPts} = 2$ and a minimum distance $\epsilon_\theta = 0.2^\circ,$ which is twice the typical PSF value,  for the space dimension and $\epsilon_t = \Delta t$ for the time dimension.
DBSCAN finds clusters by partitioning the data and eliminating isolated points. Clusters found may be spatially larger than the H.E.S.S. PSF or last longer than the chosen $\Delta t$. When this happens, photon events farthest from the cluster median position in space are excluded until the smallest enclosing circle reaches has a size of $< 0.1^\circ$. The same procedure is performed in the time dimension. The whole cluster finding procedure was checked by injecting simulated clusters of photons in the data.

\subsection{False positive background estimation}\label{sec:bckgrd}

The probability of false association of photons inside a cluster 
increases strongly with $\Delta t.$ The PBH flux also increase with $\Delta t,$ but at a much lower rate. The optimal value of $\Delta t$ comes from a compromise between a large enough PBH emission and a low background. This paper presents preliminary results for $\Delta t = 10 \mathrm{s}.$

\begin{figure}[!h]
\centering
\begin{subfigure}{.49\textwidth}
\centering
\includegraphics[clip,width=1.\linewidth]{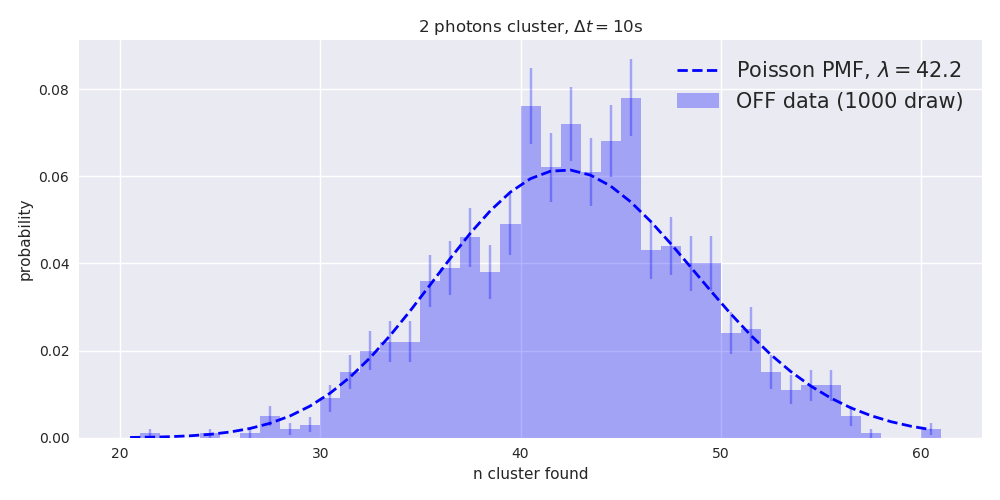}
\caption{}
\label{1gmes}
\end{subfigure}\hfill%
\begin{subfigure}{.49\textwidth}
  \centering
  \includegraphics[clip,width=.9\linewidth]{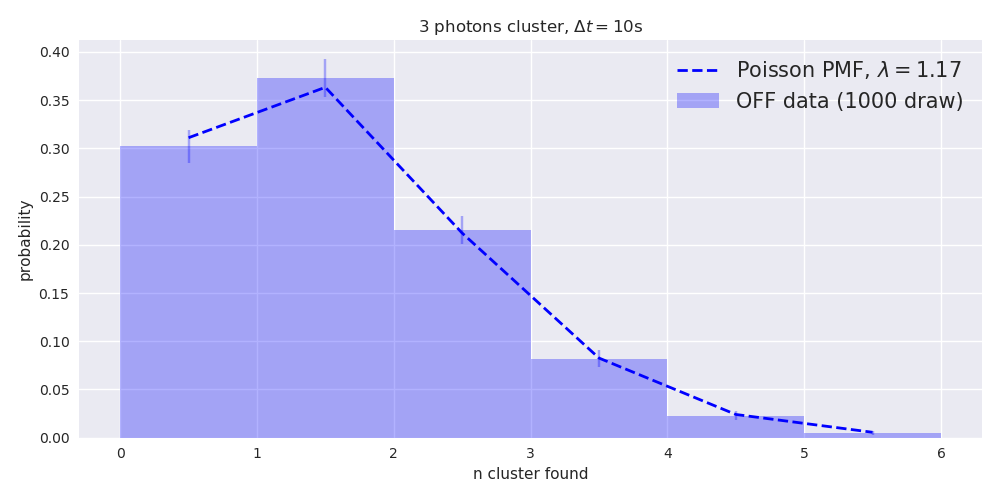}
  \caption{}
  \label{2gmes}
\end{subfigure}
\caption{Distribution of the number of clusters found in 1000 random drawings, for the cases of 2 (a) and 3 (b) photon clusters. A Poisson law (dashed line) with a mean equal to the measured mean in the OFF data is superposed.}
\label{poisson}
\end{figure}

The false positive background can be estimated directly from the data by using the same photon list with randomized ("scrambled") times of arrival. The average value of the cluster distribution obtained by time scrambling 50 times the photon list of each run is taken as the background. 
Hereafter, this set of randomized photons will be referred to as OFF data. For a given number $b$ of photon in the cluster, the number of cluster found in the OFF data follows a Poisson distribution (see fig \ref{poisson}). 

In the next section, the observation of photon clusters in the HEGS data-set of H.E.S.S is compared the expectation from the scrambled background and potential excesses are searched.

\section{Results}\label{sec:results}

The data-set of 6240 runs, totalizing 2859.1 hours of data was analyzed with the method described in sec. \ref{sec:analysis}. 

The PBH evaporation density is estimated by maximizing a likelihood ratio with $\rho_\mathrm{PBH}$ as a free parameter, following the procedure of Feldman-Cousins \cite{FeldmanCousins}. The likelihood ratio is given by:

\begin{equation}
\frac{ { \cal L}_{H_1}} { { \cal L}_{H_0}} = 
\prod_{n_\mathrm{ON} \in \mathrm{Data}} 
\frac{ { \cal P }  ( n_\mathrm{ON} | \lambda = n_\mathrm{OFF} +n_{sig}(b,\mathrm{\Delta}t, \dot{\rho}_\mathrm{PBH} )  ) }
{ { \cal P }  (n_\mathrm{ON} | \lambda = n_\mathrm{OFF} ) }
\label{eq:Like_cousin}
\end{equation}

where ${ \cal P } $ is the Poisson probability,
$n_{sig}(b,\mathrm{\Delta}t,\rho_\mathrm{PBH})$ is defined in eq. \ref{eq:ngammath}, $n_\mathrm{ON}$ is the number of clusters found in the data and $n_\mathrm{OFF}$ is the corresponding mean number of clusters found in the OFF data.

The corresponding test statistics is given by:

\begin{equation}
TS =  -2 ln \left(  \frac{ { \cal L} _{H_1}}{ { \cal L}_{H_0}} \right)  =  
2 \times \sum_{n_\mathrm{ON}} n_{sig} + n_\mathrm{ON} \left( ln(n_\mathrm{OFF}) - ln ( n_\mathrm{OFF} + n_{sig}) \right)
\end{equation}

\begin{figure}[htb]
\centering{\includegraphics[scale=0.5, clip=true]{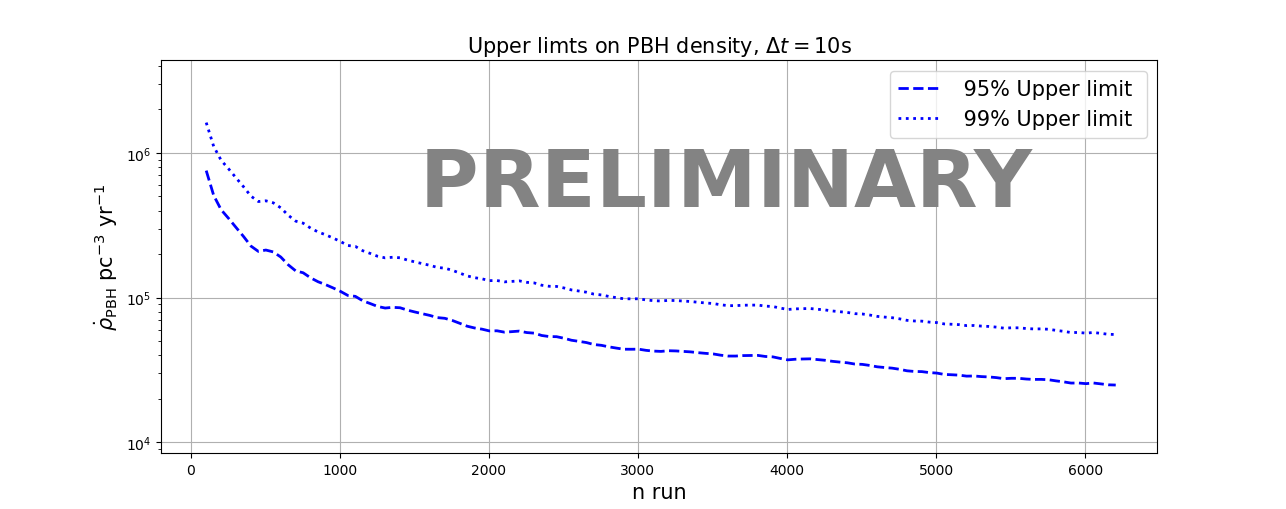}}
\caption{Evolution of the 95\% (dashed line) and 99\% (dotted line) CL upper limits on the PBH evaporation density  $\dot{\rho}_\mathrm{PBH}$ as a function of the number of runs.}
\label{fig:ulevo}
\end{figure}

$TS(\dot{\rho}_\mathrm{PBH})$ has a maximum value of $0.006,$ therefore there is no significant $\dot{\rho}_\mathrm{PBH}$ excess in the data.Upper limits with a confidence level (CL) of 95\% and 99\% can be set by finding the $\dot{\rho}_\mathrm{PBH}$ for which TS= 4 and 9 respectively. The evolution of these upper limits as a function of the number of processed run is shown on Fig.\ref{fig:ulevo}. Using the full data-set, the preliminary upper limits on $\dot{\rho}_\mathrm{PBH}$ are :

\begin{equation}
\dot{\rho}_\mathrm{PBH} < 2.5 \times 10^4 \hspace{.2cm}  \mathrm{pc}^{-3} \mathrm{yr}^{-1}   \hspace{1cm}(95\% CL)
\end{equation}
\begin{equation}
\dot{\rho}_\mathrm{PBH} < 5.6 \times 10^4 \hspace{.2cm} \mathrm{pc}^{-3} \mathrm{yr}^{-1}   \hspace{1cm}(99\% CL)
\end{equation}

\begin{figure}[htb]
\centering{\includegraphics[scale=0.5, clip=true]{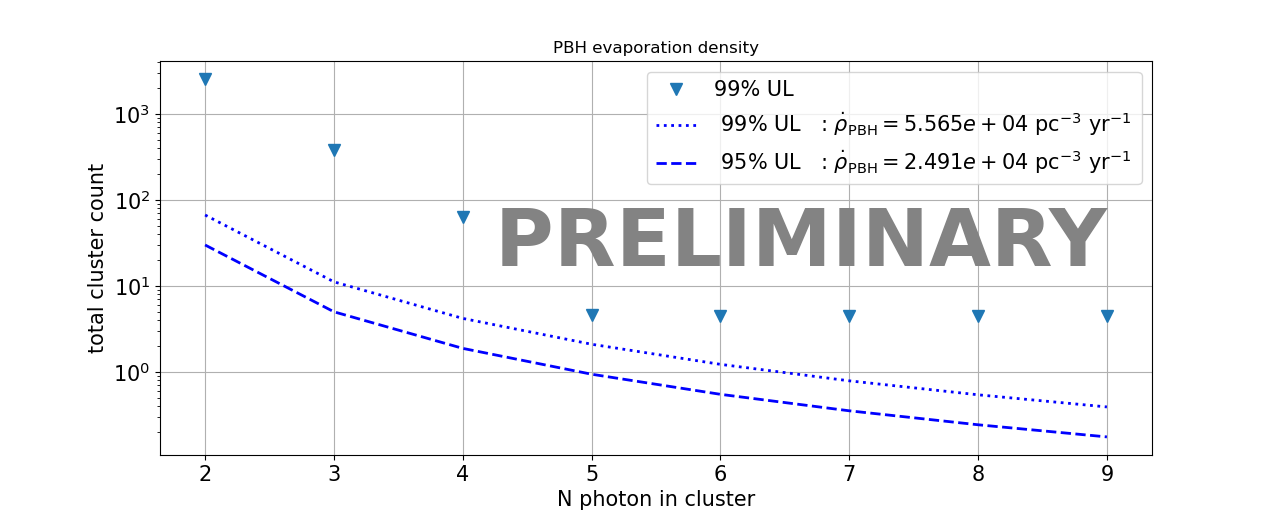}}
\caption{Expected cluster counts in the H.E.S.S. data-set for a  $\dot{\rho}_\mathrm{PBH}$  corresponding to the 95\% (dashed line) and 99\% (dotted line) CL upper limits. Blue triangles represent the model-independent upper limits on the number of clusters seen in the data.} 
\label{fig:sumary}
\end{figure}

Figure \ref{fig:sumary} shows the expected cluster distribution for the $\dot{\rho}_\mathrm{PBH}$  limit densities.
The same figure presents the limits obtained independently for each cluster size.

\subsection{Conclusion}\label{sec:conclusion}

Short time-scale photon clusters were searched in 2860 hours of H.E.S.S.
extragalactic observations. These photon clusters is a potential signature for PBH evaporation. No significant excess was found in the data. The derived preliminary 95\% CL upper limit on the PBH evaporation density : $\dot{\rho}_\mathrm{PBH} < 2.5 \times 10^4 \hspace{.05cm}  \mathrm{pc}^{-3} \mathrm{yr}^{-1}$ is competitive with previous measurements. This preliminary analysis used a  $\Delta t=10$ seconds and will be extended to smaller and larger time scales.

\acknowledgments

{\small The support of the Namibian authorities and of the University of Namibia in facilitating the construction and operation of H.E.S.S. is gratefully acknowledged, as is the support by the German Ministry for Education and Research (BMBF), the Max Planck Society, the German Research Foundation (DFG), the Helmholtz Association, the Alexander von Humboldt Foundation, the French Ministry of Higher Education, Research and Innovation, the Centre National de la Recherche Scientifique (CNRS/IN2P3 and CNRS/INSU), the Commissariat \`a l'\'energie atomique et aux \'energies alternatives (CEA),the U.K. Science and Technology Facilities Council (STFC), the Knut and Alice Wallenberg Foundation, the National Science Centre, Poland grant no. 2016/22/M/ST9/00382, the South African Department of Science and Technology and National Research Foundation, the University of Namibia, the National Commission on Research, Science \& Technology of Namibia (NCRST), the Austrian Federal Ministry of Education, Science and Research and the Austrian Science Fund (FWF), the Australian Research Council (ARC), the Japan Society for the Promotion of Science and by the University of Amsterdam.

We appreciate the excellent work of the technical support staff in Berlin, Zeuthen, Heidelberg, Palaiseau, Paris, Saclay, Tubingen and in Namibia in the construction and operation of the equipment. This work benefitted from services provided by the H.E.S.S. Virtual Organisation, supported by the national resource providers of the EGI Federation.}

\end{document}